\documentclass[]{spie}  

\usepackage{amsmath,amsfonts,amssymb}
\usepackage{graphicx}
\usepackage[colorlinks=true, allcolors=blue]{hyperref}

\title{A New Frontier for J-band Interferometry: Dual-band NIR Interferometry with MIRC-X}

\author[a]{Aaron Labdon}
\author[b]{John D. Monnier}
\author[a]{Stefan Kraus}
\author[c]{Jean-Baptiste Le Bouquin}
\author[b]{Benjamin R. Setterholm}
\author[d]{Narsireddy Anugu}
\author[e]{Theo ten Brummelaar}
\author[f]{Cyprien Lanthermann}
\author[a]{Claire L. Davies}
\author[b]{Jacob Ennis}
\author[b]{Tyler Gardener}
\author[e]{Gail H. Schaefer}
\author[e]{Lazlo Sturmann}
\author[e]{Judit Sturmann}

\affil[a]{School of Physics and Astronomy, University of Exeter, Exeter, Stocker Road, EX4 4QL, UK}
\affil[b]{Astronomy Department, University of Michigan, Ann Arbor, MI 48109, USA}
\affil[c]{Institut de Planetologie et dAstrophysique de Grenoble, Grenoble 38058, France}
\affil[d]{Steward Observatory, Department of Astronomy, University of Arizona, Tucson, USA}
\affil[e]{The CHARA Array of Georgia State University, Mount Wilson Observatory, Mount Wilson, CA 91203, USA}
\affil[f]{Instituut voor Sterrenkunde, KU Leuven, Celestijnenlaan 200D, 3001 Leuven, Belgium}

\authorinfo{Further author information: (Send correspondence to A.L.)\\A.L: E-mail: al612@exeter.ac.uk\\}

\begin{document} 
\maketitle

\begin{abstract}
In this contribution we report on our work to increase the spectral range of the Michigan Infrared Combiner-eXeter (MIRC-X) instrument at the CHARA array to allow for dual H and J band interferometric observations. We comment on the key science drivers behind this project and the methods of characterisation and correction of instrumental birefringence and dispersion. In addition, we report on the first results from on-sky commissioning in November 2019.
\end{abstract}


\section{INTRODUCTION}
\label{sec:intro}  

The Michigan Infrared Combiner (MIRC) was a six-telescope infrared beam combiner at the CHARA telescope array \cite{Monnier04}, the world’s largest baseline interferometer in the optical/infrared, located at the Mount Wilson Observatory in California. In the summers of 2017 and 2018 MIRC underwent significant upgrades, with the newly commissioned instrument being named MIRC-X. The upgrade of MIRC-X occurred in two phases, the first of these phases involved the implementation of an ultra-fast, ultra low read noise near-infrared camera \cite{Anugu18} and an overhaul of the MIRC control software. The second phase involved the replacement of the optical fibres and beam combination element in addition to the commissioning of polarisation controllers. A detailed description of the upgraded MIRC-X instrument can be found elsewhere\cite{Kraus18,Anugu20}.

Over the past 15 years MIRC/MIRC-X has produced many outstanding results in a variety of areas of astrophysics including the imaging of stellar surfaces of rapid rotators, eclipsing binary systems and star spots on the surfaces of highly magnetically active stars. \cite{Monnier12,Roettenbacher17,Schaefer19,Chiavassa20}. In more recent years, the upgrades have allowed for observations of young stellar objects (YSO) with all six telescopes for the first time \cite{Kraus20,Labdon20} and for high precision single-field astrometry for the detection of Hot Jupiter objects in binary systems \cite{Gardner20}.

Historically MIRC-X has only operated on the H band, however, one of the ultimate aims of the upgrades was to conduct dual H and J band observations. The J band ($1.1-1.4\,\mu m$) is a relatively untapped resource in long-baseline interferometry, but one with huge potential for scientific discoveries. It allows access to the photosphere in giant and super-giant stars relatively free from opacities of molecular bands. In addition, the J band traces the warmest and smallest scales of protoplanetary disks, where accretion and viscosity processes take place. Finally, the J band allows for higher resolution observations than near and mid-IR allowing us to probe the smallest scales of astrophysical objects.
 
In this paper we present a summary of the work done to commission the J band for observations at the MIRC-X instrument. In section \ref{sec:hardware} we discuss the J band specific hardware improvements required, while in section \ref{sec:software} we summarise the corresponding software developments undertaken in this endeavour. In section \ref{sec:results} we discuss the first science results obtained in this new mode. Finally in section \ref{sec:summary} we provide our concluding remarks and provide a roadmap for bringing the simultaneous H and J band observing mode into standard use at the CHARA array.

\section{Hardware requirements} \label{sec:hardware}

Beyond the major instrument upgrades in 2017 and 2018 mentioned in the introduction, various other hardware requirements were needed in order to conduct J band observations. In this section we discuss this work and the design decisions involved.

\subsection{Filters}

Operating in standard observing mode MIRC-X employs an H band filter with central wavelength $\lambda = 1.61\,\mu m$ and bandwidth $\Delta\lambda = 0.27\,\mu m$. However, in order to observe in the J band a new filter which extends to a lower wavelength regime must be employed. However, this is hampered owing to the presence of a metrology laser at CHARA. This laser operates at $1.319\,\mu m$ and is responsible for providing accurate metrology of the delay line cart position, a vital role at any optical interferometer. This laser wavelength lies directly between the H and J bands and will thus contaminate any stellar flux on the detector. 

\begin{figure} [ht]
   \begin{center}
\begin{tabular}{c} 
   \includegraphics[height=7cm]{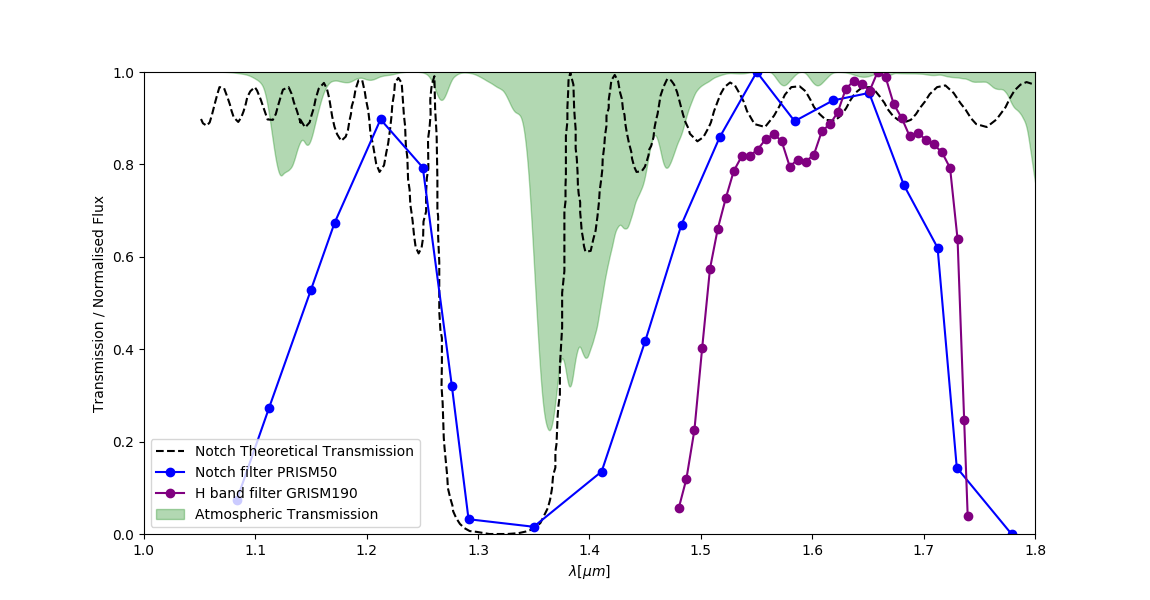}
   \end{tabular}
   \end{center}
   \caption[example] 
   { \label{fig:Atmosphere} 
Theoretical and measured transmission of the notch filter across the H and J bands. The dashed black curve represents the theoretical transmission curve of the filter. The solid green area is the atmospheric transmission. The blue line with points is the normalised flux measured using the MIRC-X detector and the CHARA STS through the notch filter and the PRISM50. The solid purple line with points is the normalised flux measured using the MIRC-X detector and the CHARA STS through the H band filter and the GRISM190, the current highest spectral mode of MIRC-X.   }
   \end{figure}

This was overcome with addition of a notch or band-stop filter. These filters reject/attenuate signals in a specific wavelength band and pass the signals above and below this band. In this way we can reject light from the CHARA metrology laser and accept stellar contributions from the H band J bands above and below this respectively. The theoretical and measured characteristics of the notch filter we employed is shown in Figure\,\ref{fig:Atmosphere}.

\subsection{Longitudinal Dispersion Compensators}

Long-baseline optical interferometry poses unique engineering challenges. One of the most demanding aspects is the requirement for optical path length compensation. The path length of the various light beams are not identical for each telescope, owing to the position of the scope and position of the astronomical object on the sky. A such, the path difference much be corrected in order to ensure coherence for interferometric fringes. At CHARA this is done using two methods, the first is to employ fixed amounts of delay which can be moved in and out of the light path using retractable mirrors. These are known as Pipes of Pan or PoPs. The second and universally used method is to employ carts affixed to delay lines which track the geometric phase difference as a stars moves across the sky. The difficulty with the second method is that it very difficult to achieve in a vacuum, due to the size and number of moving parts. This introduces a large amount of air into the system, potentially up to $80\,m$ difference in air path between the beams. This air leads to atmospheric, chromatic dispersion. 

In standard single band observing, this does not pose a problem due to the small wavelength range. However, in dual band observations this become a major problem. If delay line carts are tracking H band fringes, the J band light will be incoherent due to chromatic dispersion and fringes will be severely reduced contrast to the point of being un-observable. Therefore a method of correcting for atmospheric dispersion must be employed. At CHARA, this is in the form of Longitudinal Dispersion Compensators (LDCs). 

\begin{figure} [ht]
   \begin{center}
\begin{tabular}{c} 
   \includegraphics[height=6.2cm]{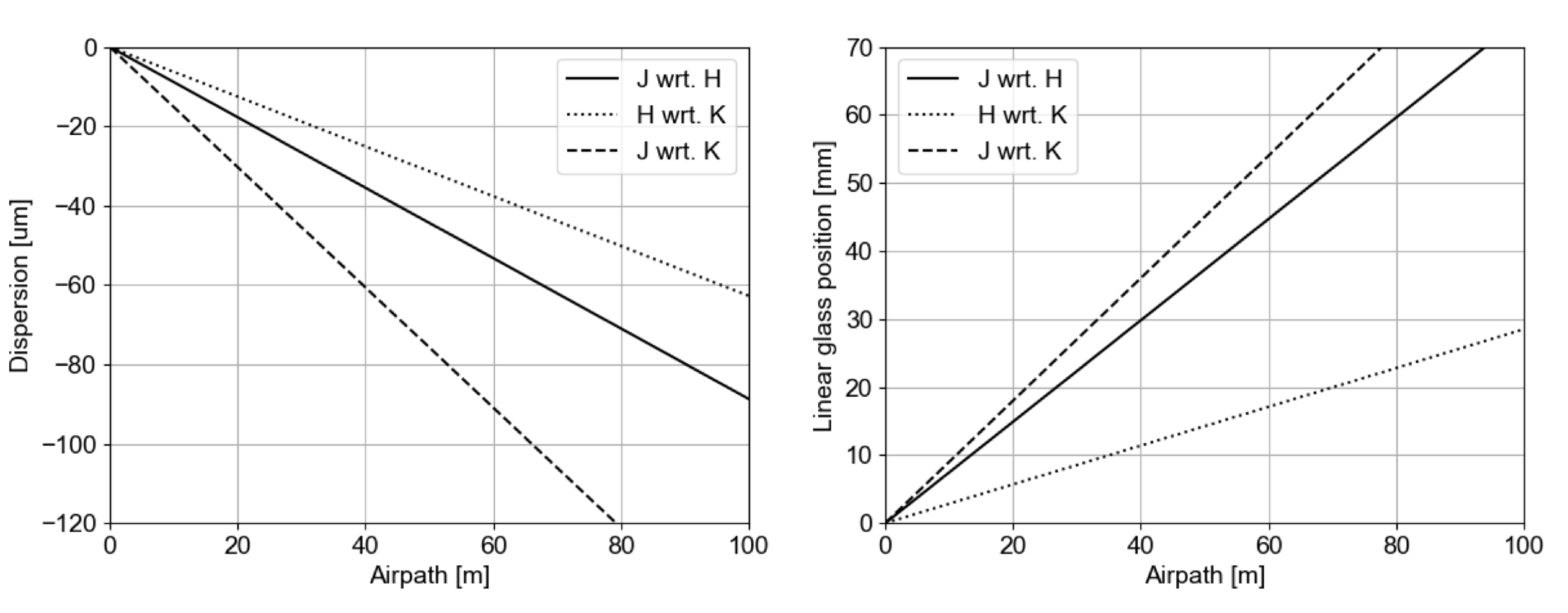}
   \end{tabular}
   \end{center}
   \caption[example] 
   { \label{fig:AirDisp} 
LEFT: Theoretical atmospheric dispersion between the NIR bands covered by MIRC-X and the upcoming MYSTIC instrument. Group delay measurements are used an analogy for chromatic dispersion. The J band is taken as $1.2\,\mu m$, the H band as $1.55\,\mu m$ and the K band as $2.2\,\mu m$. The solid line represents the dispersion in the J band with respect to the H, the dotted line the dispersion in H with respect to k and the dashed the dispersion in J with respect to K. RIGHT: The linear LDC glass movement required to correct the dispersion between the J, H and K bands. The wavelengths adopted and line meaning are the same as shown on the left.  }
   \end{figure}

The LDCs consist of wedges of glass that can move in and out of the beam to increase or decrease the thickness of glass in the beam path. The characteristics of the LDCs are described in great detail in\cite{Berger03} when the LDCs were commissioned in order to enable simultaneous near-infrared (NIR) observations with the CLIMB instrument and visible observations with PAVO\cite{Ireland08}. Despite not being designed for use between the H and J bands, the choice of Schott-10 glass proved ideal for use with MIRC-X. Shown in Figure\,\ref{fig:AirDisp} is the expected dispersion between the NIR wavebands and the expected linear movement of the glass to successfully correct for this. The thickness of glass required for a given air path can be calculated using
\begin{equation}
\label{eq:glass_per_m}
\Delta L = \Delta x \times \frac{n_a(\sigma_1) - n_a(\sigma_2) + \sigma_1 n_a^{'}(\sigma_1) - \sigma_2 n_a^{'}(\sigma_2)} {n_g(\sigma_1) - n_g(\sigma_2) + \sigma_1 n_g^{'}(\sigma_1) - \sigma_2 n_g^{'}(\sigma_2)} \, .
\end{equation}
Here, $\Delta L$ is the thickness of glass, $\Delta x$ is the air path in meters, $n_a$ is the refractive index of air and $n_g$ is the refractive index of the LDC glass. $\sigma_1$ and $\sigma_2$ refer to the wavenumbers of the two observing wavelengths. In our model the refractive index of air was calculated using a composite of Ciddor\cite{Ciddor96} and Mather\cite{Mathar07} models in order to cover the full NIR. Atmospheric conditions, such as temperature and humidity were taken to be average values for the CHARA laboratory. More information on the dependence of the refractive index of air on these parameters can be found elsewhere \cite{Ciddor96,Mathar07}.

Overall the LDCs are capable of correcting for atmospheric longitudinal dispersion between the H and J bands, making simultaneous observations possible.

\section{Software control and Data Reduction} \label{sec:software}

A variety of software developments were required in order to conduct these observations. In addition to changes to the real-time operating code of MIRC-X/CHARA, the reduction pipeline had to be modified, as outlined below. Also, we independently verified the data products by calibrating known calibrator stars and ensuring the stability of the transfer function.

\begin{figure} [ht]
   \begin{center}
\begin{tabular}{c} 
   \includegraphics[height=7cm]{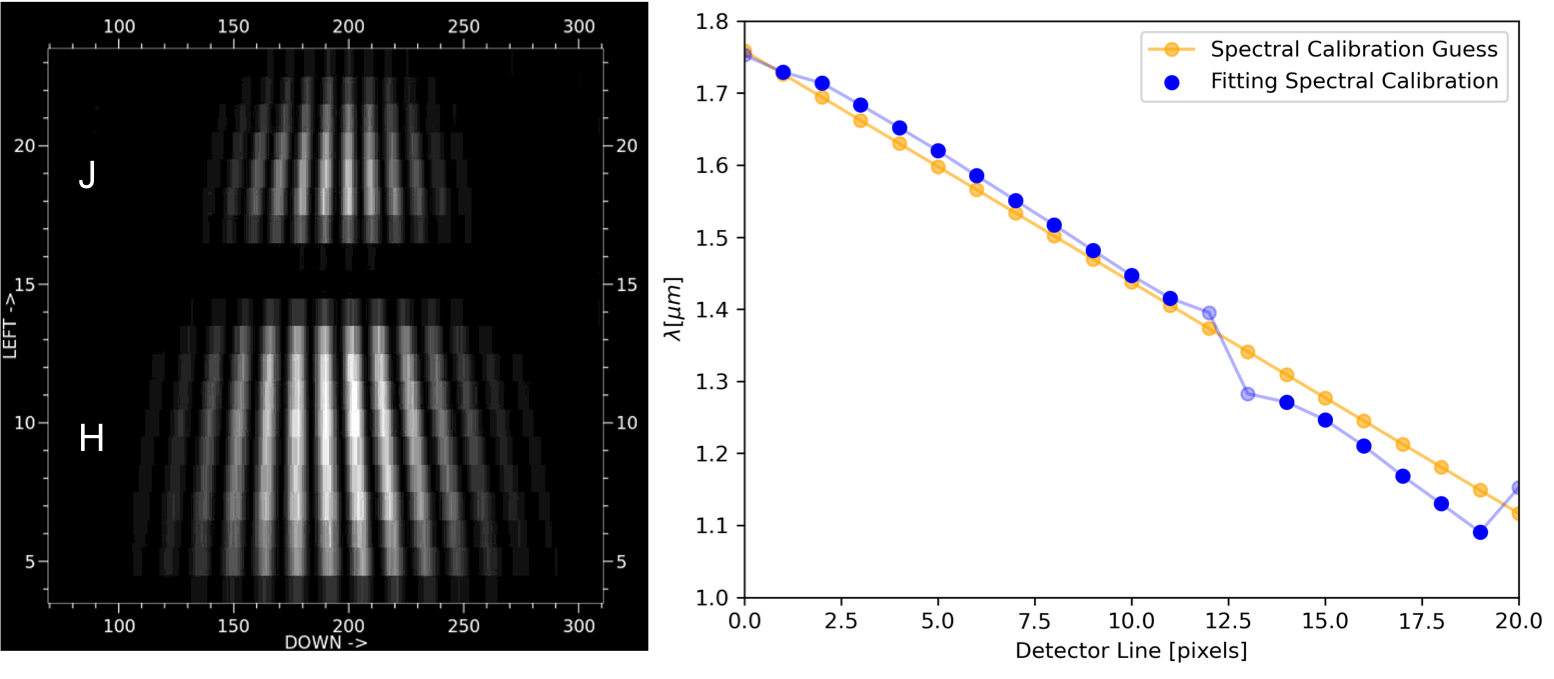}
   \end{tabular}
   \end{center}
   \caption[example] 
   { \label{fig:speccal} 
LEFT: MIRC-X detector window showing laboratory induced fringes along a single baseline. The wavelength increases towards the bottom, meaning the J band is on top. The notch filter rejection window is clearly visible between the H and J bands at $\sim1.3\,\mu m$. RIGHT: Spectral calibration of the MIRC-X simulated data. The inital guess of the spectral calibration is shown in pale orange and the fitted calibration is shown in blue. The translucent blue points represent those that were rejected by the pipeline due to low flux.}
   \end{figure}
   
\subsection{LDC Control Software}

The control of the LDCs is based upon that developed by Berger et al.\cite{Berger03} for controlling dispersion between the visible and NIR. This software was adapted and expanded by adopting the dispersion models described in section\,\ref{sec:hardware}. The basic function of the software is to receive the delay line cart positions in order to calculate the total amount of atmosphere in the system. Using equation\,\ref{eq:glass_per_m} it  calculates the thickness of glass required to correct for this dispersion, which is then translated into the linear movement for the LDC. This is done at a rate of one LDC per second, meaning a full movement cycle of the LDCs is 6 seconds. This is sufficient given the slow rate of change of the total air path length in the system.

\subsection{Data Reduction Pipeline}

The data reduction pipeline for MIRC-X is already well established and publicly available\cite{Anugu20}. It has a proven track record through several publications. However, it was not designed to reduce J band data and so several changes had to be made to the existing code. 

The first change to the data reduction pipeline consisted of adjusting the fringes and photometry windows to accurately detect and account for the J band flux. Previously the flux window had been detected by fitting a simple Gaussian to the detector flux, however, due to the presence of the notch filter the simultaneous data is double-peaked. Figure\,\ref{fig:speccal} shows the detector window of MIRC-X covering both the H and J bands, the spectral range of the J band is just over half that of the H band.


The second change comes from the spectral calibration, that is the process of determining the wavelength of each spectral channel. The wavelength is assigned by fitting the frequency of the fringe patterns, based on an initial guess. The inital guess is based on the known spectral dispersion of the prism used, in this case $R = 50$. Figure\,\ref{fig:speccal} shows the spectral calibration result obtained from the MIRC-X data reduction pipeline for artificially produced interference fringes. Both the inital guess and the final fitted spectral calibration are shown.  

\begin{figure} [ht]
   \begin{center}
\begin{tabular}{c} 
   \includegraphics[height=12cm]{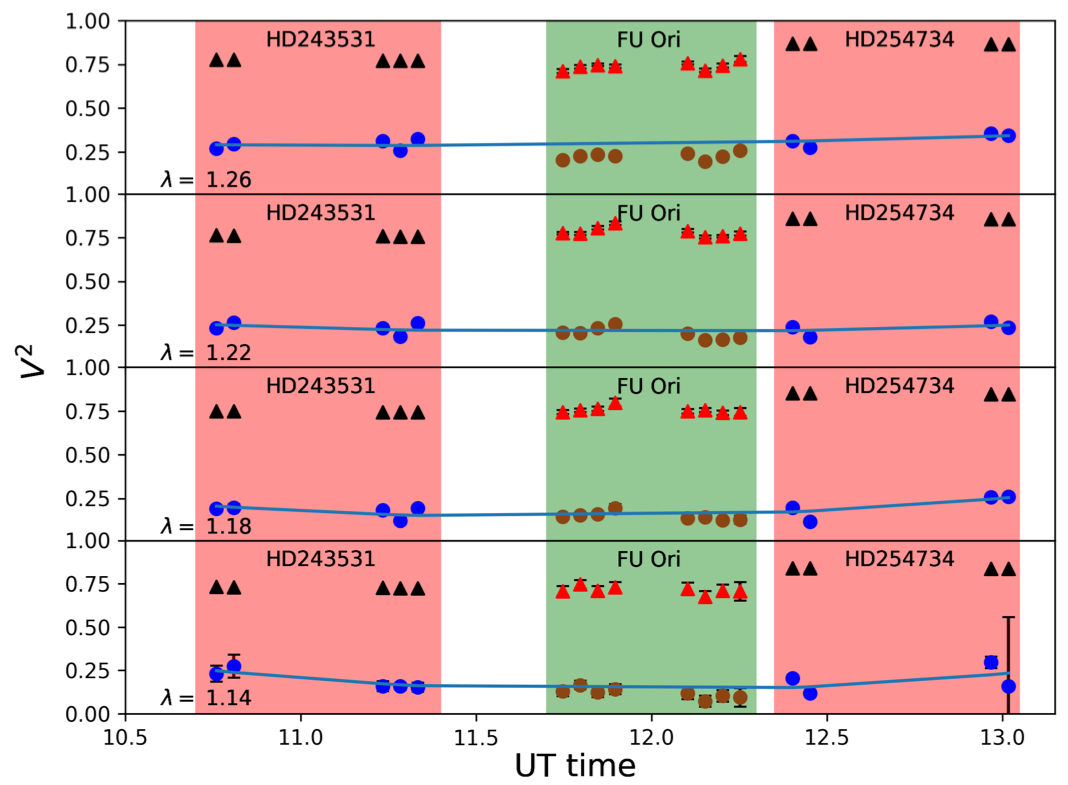}
   \end{tabular}
   \end{center}
   \caption[example] 
   { \label{fig:TFucntion} 
Calibration transfer function for 2019 November 19. The red-shaded areas are those representing calibrators and green are those representing science observations of FU\,Ori \cite{Labdon20}. these are along a single baseline (S1-S2) for the four J band spectral channels. The blue and brown points are the raw, uncalibrated squared-visibilities for the calibrators and science target respectively. The black and red triangles are the calibrated squared-visibilities for the calibrators and science target respectively. the blue represents the transfer function across the calibrators. }
   \end{figure}

\subsection{Data Calibration}

In order to ensure the validity of the data products, careful calibration tests were undertaken. The calibration procedure was unchanged from the standard MIRC-X pipeline as described in great detail elsewhere\cite{Anugu20}. Figure\,\ref{fig:TFucntion} shows the transfer function on a single baseline across the three hour observing period on the night of 2019 November 19. The 4 wavelength channels are those across the J band only, as the calibration of the H band channels has already been verified in previous papers. The linearity of the transfer function shows the stability of the observing across the 2.5 hour period. Large jumps in the transfer function on short timescales would indicate instability/inaccuracies in the dispersion control subsystems.

\section{First Science Results} \label{sec:results}

Following the upgrades to the hardware and software described in this paper, we conducted the first science observations in November 2019. We adopted a 4-telescope combination using the first four fibres (lowest spatial frequencies) of MIRC-X, this is the only possible observing mode with the J band as the highest spatial frequencies are undersampled (sub-Nyquist). As the first science target, we observed the young outbursting object FU\,Orionis. The data was reduced following the reduction and calibration steps described in section\,\ref{sec:software}. Full details of the data analysis and conclusions can be found in\cite{Labdon20}.

\begin{figure} [ht]
   \begin{center}
\begin{tabular}{c} 
   \includegraphics[height=7cm]{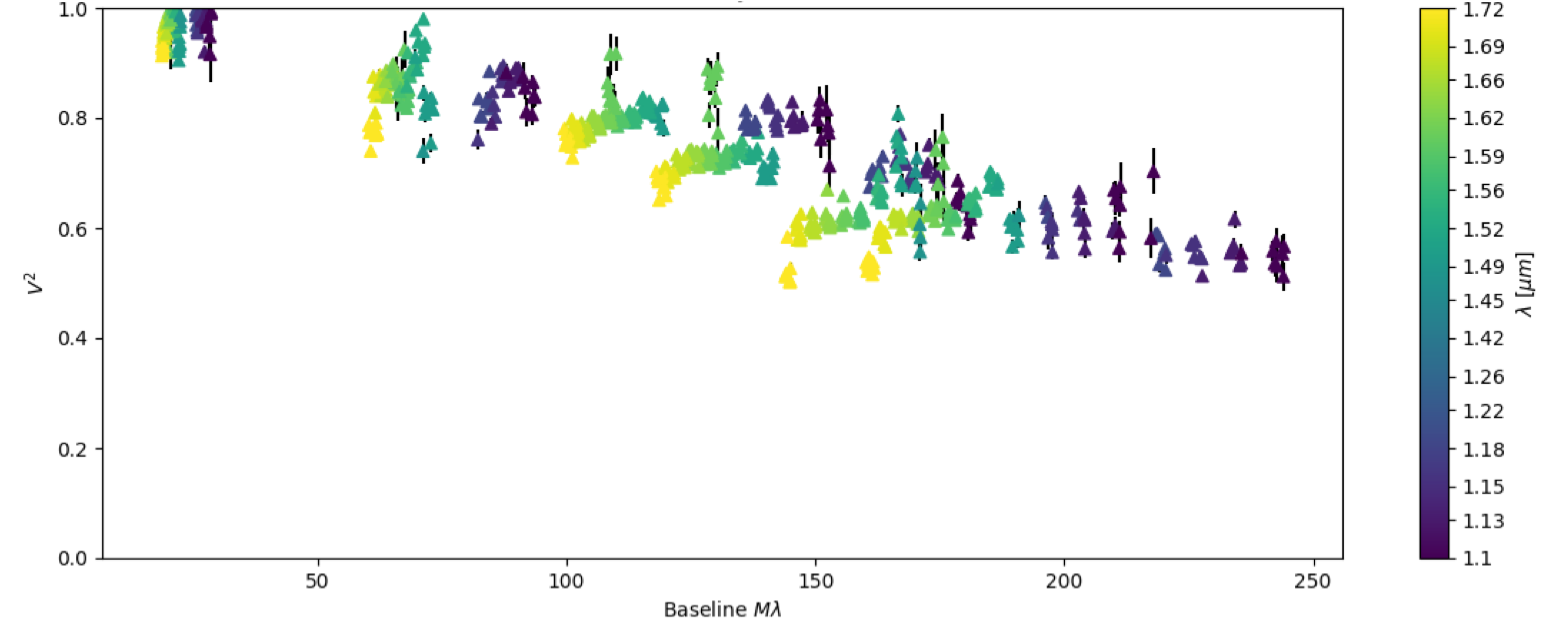}
   \end{tabular}
   \end{center}
   \caption[example] 
   { \label{fig:FUOri_vis} 
Squared visibilites for the object FU Orionis with the MIRC-X instrument. Colour scale represents the wavelength of the observations. Dark-blue/purple points are across the J band, while green-yellow ponts are across the H band.}
   \end{figure}

We obtained two J+H band pointings on FU\,Ori with MIRC-X in 2019, using a 4-telescope configuration in CAL-SCI concatenation sequences. Of the available baselines, a maximum physical baseline of $280\,$m was used corresponding to a maximum resolution of $0.34\,\mathrm{mas}$. The calibrated squared visibilities are shown in Figure\,\ref{fig:FUOri_vis}. These observations represent the first J band interferometric observations of a young stellar object and allowed us to probe the innermost regions of this highly accreting disk. In addition to constraining the 2-D geometry, these multi-wavelength observations allowed us to probe the temperature gradient profile of the inner disk. Theoretical works have proposed models for the expected temperature profile of accretion disks since the late 1970's\cite{Shakura73,Pringle72,Pringle81}. These models follow the simple prescription
\begin{equation}
\label{eq:temp}
T_{R} = T_{0}(R/R_{0})^{-Q} \, ,
\end{equation}
where $T_0$ is the temperature at radius $R_0$ and $Q$ is gradient exponent. The two models proposed in the literature provide alternate $Q$ parameters. For isothermal flared discs, heated only by reprocessed stellar radiation, a value of $Q = 0.5$ is predicted\cite{Kenyon87,Dullemond04}.  Actively accreting, steady-state disks that are viscously heated should exhibit a value of $Q = 0.75$ \cite{Pringle81}. 

\begin{figure} [ht]
   \begin{center}
\begin{tabular}{c} 
   \includegraphics[height=8cm]{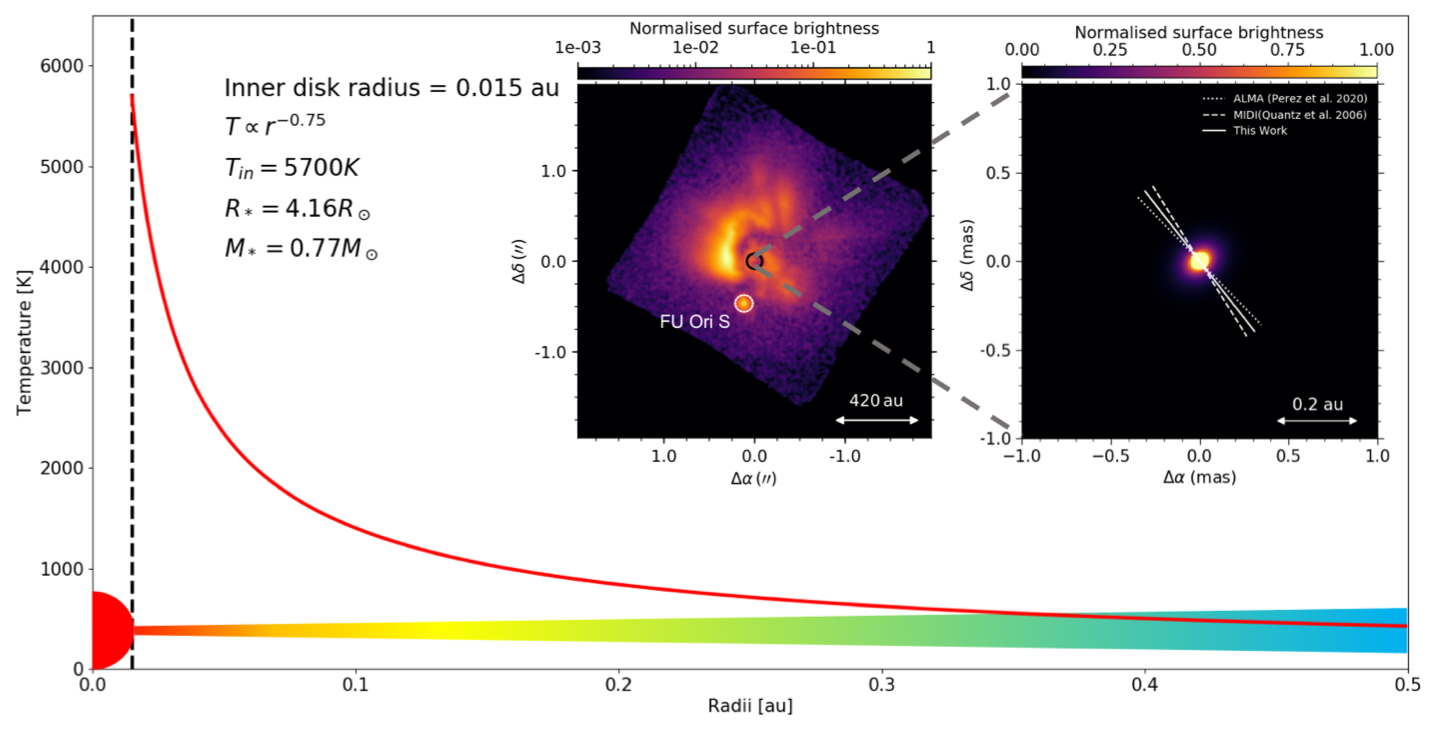}
   \end{tabular}
   \end{center}
   \caption[example] 
   { \label{fig:FUOri_results} 
Temperature gradient of FU\,Ori over the inner disk. The temperature and location of the inner radius of the disk are $5800\,\mathrm{K}$ and $0.015\,\mathrm{au,}$ respectively. The dashed line represents the inner disk edge. The vertical profile of the disk schematic is not to scale, nor does it represent the vertical temperature structure of the disk. {\it INSET-left:} Linearly polarized intensity, scattered light image obtained with the Gemini Planet Imager (GPI) in the J-band \cite{Laws20}. The image is centered on FU\,Ori\,N within the black circle. FU\,Ori\,S is also visible to the South-West. {\it INSET -right:} Brightness distribution of the TGM disk model in the J-band, as derived in this work. In both panels, we plot normalised surface brightness with a logarithmic colour scale. The white lines in the image indicate the position angle of the disk minor-axis derived from other observations in literature and in this work.}
   \end{figure}

For FU Orionis we measured $Q = 0.74\pm0.02$, a value that indicates  viscously heated material. Our analysis also revealed that the disk inner radius is equivalent to the stellar radius, meaning the disk extends down to the stellar surface. This is indicative of boundary layer accretion dominating over magnetospheric accretion scenarios. A summary of our results is shown in Figure\,\ref{fig:FUOri_results}. This is among the first direct interferometric evidence for both boundary layer accretion and viscously heated inner accretion disks.  

\section{Summary and Future Work} \label{sec:summary}

\subsection{Conclusions}

In conclusion, we present a new and unique observing mode for the MIRC-X instrument. The simultaneous H and J band mode provides an exciting opportunity to conduct very high resolution observations at a little explored wavelength in interferometry. The J band wavelength regime is highly interesting scientifically across a range of astrophysical objects from stellar photospheres for circumstellar accretion disks.

We outlined the work to bring the J band to MIRC-X, including the inital hardware and software development and later the work done to enable data reduction and calibration in a reliable way.

We also present the first science results from the endeavour in the form of a detailed study of the accretion disk around FU\,Orionis. We find evidence of boundary layer accretion and viscous heating in this first-of-its-kind study. In the process we demonstrate the effectiveness and application of this new observing mode as it opens up a new frontier in interferometry. 

For more information regarding MIRC-X observing modes and characterisation see other works in these proceedings by Setterholm et al.\cite{Setterholm2020} and Anugu et al.\cite{Anugu20}.

\subsection{Future Work}

The future of the J band at MIRC-X lies in community access. Follow-up observations in this observing mode are expected in late 2020 where we expect to perfect the observing routine and confirm dispersion models on sky. Following this, we aim to release the J band to the wider community and offer it as a standard MIRC-X mode from semester 2021-B onwards. 

Furthermore, the J-band also represents a promising wavelength regime for the VLT Interferometer, where the J-band visitor instrument BIFROST could enable high-spectral resolution observations in spectral lines and open new science applications on star formation, fundamental stellar astrophysics, and the characterisation of star and planetary systems  \cite{kraus19}.



\acknowledgments 
 
We acknowledge support from an STFC studentship (No.\ 630008203) and an European Research Council Starting Grant (Grant Agreement No.\ 639889). JDM acknowledges funding from NASA NNX09AB87G, NSF NSF-ATI 1506540, and NASA XRP Grant NNX16AD43G.
This work is based upon observations obtained with the Georgia State University Center for High Angular Resolution Astronomy Array at Mount Wilson Observatory.  The CHARA Array is supported by the National Science Foundation under Grant No. AST-1636624 and AST-1715788.  Institutional support has been provided from the GSU College of Arts and Sciences and the GSU Office of the Vice President for Research and Economic Development.
MIRC-X received funding from the European Research Council (ERC) under the European Union's Horizon 2020 research and innovation programme (Grant No.\ 639889).  

\bibliography{report} 
\bibliographystyle{spiebib} 

\end{document}